\documentclass[pra,preprint,showpacs,showkeys]{revtex4-1}
\usepackage{graphicx}
\usepackage[latin1]{inputenc}
\usepackage{multirow}

\begin{document}

\title{Spectroscopy of PTCDA attached to rare gas samples: clusters vs. bulk matrices. I.~Absorption spectroscopy}

\author{Matthieu Dvorak$^1$, Markus M\"uller$^1$, Tobias Knoblauch$^2$, Oliver B\"unermann$^{1,3}$, Alexandre Rydlo$^4$, Stefan Minniberger$^4$, Wolfgang Harbich$^4$ and Frank Stienkemeier$^1$}
\affiliation{$^1$ Physikalisches Institut, Universit\"at Freiburg, Hermann-Herder-Str.~3, 79104 Freiburg, Germany\\
$^2$ 1. Physikalisches Institut, Universit\"at Stuttgart, Pfaffenwaldring 57, 70550 Stuttgart, Germany\\
$^3$ Institut f\"ur Physikalische Chemie, Georg-August-Universit\"at, Tammannstr. 6, 37077 G\"ottingen, Germany\\
$^4$ Institut de Physique des Nanostructures, École Polytechnique Fédérale de Lausanne (EPFL), CH-1015 Lausanne, Switzerland}

\begin{abstract}

\pacs{47.55.D-,36.40.-c,33.20.Kf,67.40.Yv}

\keywords{rare gas cluster, rare gas matrix, argon, neon, hydrogen, PTCDA, helium droplets, superfluidity}

The interaction between PTCDA (3,4,9,10-perylene-tetracarboxylic-dianhydride) and rare gas or para-hydrogen samples is studied by means of laser-induced fluorescence excitation spectroscopy. The comparison between spectra of PTCDA embedded in a neon matrix and spectra attached to large neon clusters shows that these large organic molecules reside on the surface of the clusters when doped by the pick-up technique. PTCDA molecules can adopt different conformations when attached to argon, neon and para-hydrogen clusters which implies that the surface of such clusters has a well-defined structure and has not liquid or fluxional properties. Moreover, a precise analysis of the doping process of these clusters reveals that the mobility of large molecules on the cluster surface is quenched, preventing agglomeration and complex formation.
\end{abstract}

\date{\today}

\maketitle

\section{Introduction}
The structure and phase state (solid-like or liquid-like) of small aggregates have been subject to controversial discussion since the early developments of cluster physics. In spite of the huge improvements in spectroscopic techniques during the last decades, some of their fundamental characteristics still remain unclear. For rare gas clusters, the study of these characteristics is further limited by their transparency in the visible range, limiting the use of many standard spectroscopic methods. To circumvent this problem, rare gas clusters are doped with a chromophore (atom or molecule). Studying the spectral response of this chromophore as a function of the cluster parameters allows to gain insights
into the cluster characteristics. This concept has already been developed in the 1980s~\cite{Gough1985} and applied successfully since then (see for example Ref.~\cite{Jortner1992,Hartmann1995,Leutwyler1990}), being nowadays a common method.

Among the different cluster characteristics studied, conclusive information could be obtained on the cluster phase state and phase transition as a function of size. Benzene-doped argon clusters show a liquid-solid transition at a cluster size of $N=21$.\cite{hahn88} For xenon-doped neon clusters, this transition lies at $N=300$.\cite{pietrowski_neon} To our knowledge no such studies have been performed involving large doped hydrogen clusters. Large neat hydrogen clusters obtained by supersonic expansion have a temperature of 4.2 to 4.5\,K.\cite{Knuth1995} For smaller clusters, Monte-Carlo
simulation postulated a superfluid phase for clusters sizes less than 26 molecules and for temperatures below 1.5\,K.\cite{Khairallah2007} These numbers have been obtained for para-hydrogen which correspond to molecules with their two proton spins forming a singlet state. Experiments on large hydrogen clusters involving Raman spectroscopy revealed a solid structure when formed in a supersonic expansion.\cite{kuyanov08} Only the co-expansion of helium and hydrogen with less than 1\,\% hydrogen \cite{kuyanov08} or the aggregation mechanism of somewhat smaller hydrogen clusters in helium
droplets have been found to show liquid-like properties of the clusters.\cite{momose,kuma2011}

Large rare gas clusters have been shown to provide an ideal support for chemical reactions and catalysis due to the confinement of the reactants to the clusters' surface and the corresponding high reaction probabilities\cite{Mestdagh1994}. Correspondingly, single atoms as well as di- and triatomic molecules and very few larger molecules such as SiF$_4$ and SF$_6$ have been attached to large argon clusters.\cite{Gu1990,Biquard1995,Lallement1992,Lallement1993a}

Historically, the study of doped rare gas matrices was motivated by astrochemistry studies. Indeed, perylene derivatives as well as other large polycyclic aromatic hydrocarbons (PAH) have long been postulated to compose a non-negligible part of interstellar matter\cite{leger1984} and to partly explain the presence of absorption features in the visible spectral region, the so-called diffuse interstellar bands (DIB). \cite{Salama1995} The study of such molecules in weakly interacting media at low temperatures should ascertain these postulates.

\begin{figure}
{\includegraphics[width=7.5cm]{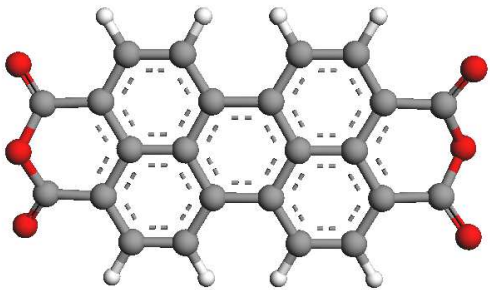}}
  \caption{The PTCDA molecule (3,4,9,10-perylene-tetracarboxylic-dianhydride, C$_{24}$H$_{8}$O$_{6}$) is composed of a perylene core and two anhydride
end-groups.}
  \label{PTCDA}
\end{figure}
In the course of the important role of the clusters' properties, several more detailed questions remain: is the cluster surface also solid? If one dopes these clusters after their formation, can one expect that the chromophores remain at the clusters' surface or can the deposited energy induce surface melting which would allow the chromophore to immerge? Provided that the chromophores remain on the clusters' surface, how is the surface mobility? In the case of multiple doping, can these chromophores aggregate and form complexes? The present work answers these questions for argon,
neon and para-hydrogen clusters. We compare spectroscopic results of PTCDA chromophores attached to large rare gas and hydrogen clusters with measurement on PTCDA embedded in rare gas matrices for which the inside localization and the solid structure are well established.

PTCDA (C$_{24}$H$_{8}$O$_{6}$, 3,4,9,10-perylene-tetracarboxylic-dianhydride, see.~Fig.~\ref{PTCDA}) is a perylene derivative which has been already studied at varies conditions. Deposited on surfaces, this planar molecule can self-assemble into well-defined orientations.\cite{Qing2009, Schreiber2000} Preferentially a herringbone structure is favored for multilayer films or bulk crystals. $\pi$-stacking of PTCDA molecules leads to semi-conducting properties of the organic layers.\cite{forrest_review,Mueller2011, Hennessy1999} These self-assembly and semi-conducting
properties are much desired for the development of efficient electro-optical devices.\cite{Peumans2003,Hains2010,Lunt2010} Therefore, PTCDA is not only studied for a fundamental understanding of organic semiconducting properties but because it is also found in device assemblies.

As far as bulk matrix isolated studies are concerned, among the many measurements obtained for ions and smaller molecules (see a list of them in Ref.~\cite{Jacox2003}), a few studies were made on larger PAHs and their ions embedded in argon and/or in neon matrices: C$_{60}$,\cite{Sassara1996} naphthalene,\cite{Salama1991} phenanthrene,\cite{salama1994} pentacene,\cite{Halasinski2000} 9,10-dichloroanthracene,\cite{crepin1990}
benzo[g,h,i]perylene,\cite{Chillier2001} and perylene.\cite{Joblin1995,Joblin1999} For PTCDA, however, only measurement in SiO$_2$ matrices are known to us.\cite{engel2006}

The present article is structured as follows: first, absorption measurements of PTCDA molecules embedded in a neon matrix are presented. In a second step, PTCDA molecules are attached to large neon, argon and para-hydrogen clusters. Comparison between matrix and clusters measurements allow decisive conclusions on the site occupation of the attached PTCDA molecules. Finally, a careful analysis of the shape of the observed optical features as well as their dependency on external parameters allow to draw conclusions on the conformation and the mobility of PTCDA molecules on such clusters.

Our related results on fluorescence emission spectroscopy of the same systems (referred to as Paper II) will be published in a separate publication.\cite{Dvorak_P2}

\section{Experimental Setup}\label{setup}
\begin{figure}
{\center\includegraphics[width=8.5cm]{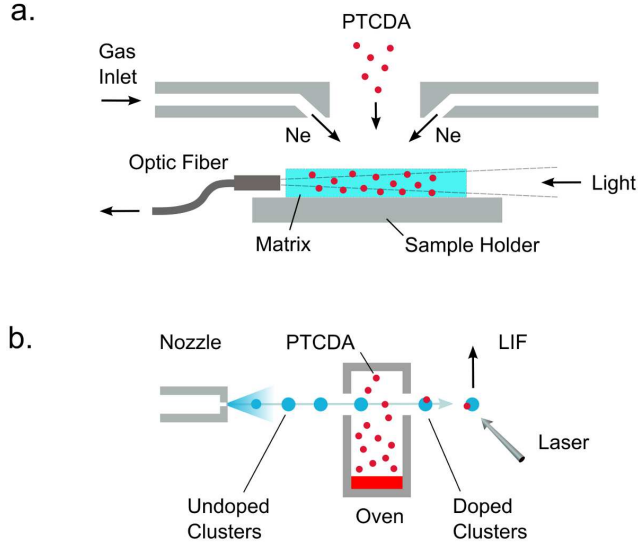}}
  \caption{Experimental setup for the matrix deposition (a) and the cluster beam spectroscopy (b).}
  \label{setup_pict}
\end{figure}
PTCDA-doped neon matrices are produced by co-deposition of neon and PTCDA on an aluminum mirror cooled by a pulse tube cryocooler (SRP-052A Cryocooler, Sumitomo Heavy Industries), see Fig.~\ref{setup_pict}\,a. An effusive beam of PTCDA molecules is emitted by an oven placed about 30\,cm away from the sample holder. The PTCDA flux is controlled by the oven temperature. Neon gas, purified by cryotrapping is co-deposited at a ratio of about $5\cdot10^4:1$ with PTCDA.
A standard 100\,W quartz tungsten halogen lamp (QTH) and a deuterium lamp (Hamamatsu L7296) cover the optical range from $10\,000-50\,000$\,cm$^{-1}$. The light is focused on a narrow slit on the side of the 50\,$\mu$m thick matrix which is used as a light guide.\cite{Conus2006} After crossing the PTCDA-rare gas matrix, the transmitted light is collected by an optical fiber (Ocean Optics Inc.~400-SR, core diameter: 400\,$\mu$m) coupled to a spectrograph (Jobin-Yvon T64000) equipped with a liquid nitrogen cooled CCD detector (Spectrum One). The total useful transmission range strongly
depends on the matrix quality and in the case of Ne ranges from $8\,000-45\,000$\,cm$^{-1}$. Further details on this setup and the applied method can be found in Ref.~\cite{Conus2006}.

Large argon, neon and para-hydrogen clusters as well as helium droplets are produced via the expansion of highly pressurized gas through a nozzle, see Fig.~\ref{setup_pict}\,b. For argon and para-hydrogen clusters, a $10\,\mu$m nozzle is used (continuous beam); for neon clusters and helium droplets a pulsed nozzle has been used (Even-Lavie valve, $60\,\mu$m trumpet nozzle, operated at 500\,Hz repetition rate).

Normal hydrogen has a natural para/ortho relative concentration of about 25\,\%/75\,\% at standard temperature and pressure. We employ an online catalytic converter to transform ortho-H$_2$ into para-H$_2$: before entering the cluster source the hydrogen gas passes through a small container cooled down to 15\,K and filled with a catalyst (here Al$_2$O$_3$ powder).\cite{I92} In this way nearly all the ortho-H$_2$ is converted into para-H$_2$ (at 20\,K and thermal equilibrium already 99.8\,\% of the molecules are in the para state.\cite{Rock1969})

The mean size of rare gas and para-hydrogen clusters depends on the expansion parameters and can be estimated with the help of scaling laws~\cite{Hagena} and literature values of measured sizes of argon,\cite{Cuvellier1991,Gaveau2000} neon,~\cite{Gaveau2000} and para-hydrogen clusters.~\cite{Obert_H2,knuth2004} With the expansion parameters used for argon (90 bars, 300\,K), the clusters have a mean size of about 450 atoms. Neon clusters (85 bar, 90\,K) consist of about 6\,000 atoms and para-hydrogen clusters (20 bar, 43\,K) have a mean size of roughly 14\,000 molecules.

Helium nanodroplets produced via supersonic expansion (60~bar, 20\,K, mean size: 20\,000 atoms) are used as reference because their properties are well known.\cite{ToeVil} The dopants --- with the exception of alkaline and some alkaline earth atoms --- are known to be located at the center of helium droplets and thermalize with the droplets temperature (0.37\,K).\cite{ToeVil} This low temperature leads to a great simplification of vibronic spectra as only the lowest vibrational level of the electronic ground state is populated. Moreover, the weak interaction with the surrounding
helium leads to only minor perturbations and narrow lines. Excitation spectra of PTCDA molecules embedded in helium droplets have been extensively discussed in Ref.~\cite{W_mono}.

After formation, the clusters are doped by means of the pickup technique in a heated oven cell (length 12\,mm) containing PTCDA powder (Sigma Aldrich, used without further purification). By varying the oven temperature, the PTCDA partial pressure and hence the mean number of PTCDA molecules picked-up can be changed. In a separate vacuum chamber the cluster beam intersects the laser beam used for spectroscopy. The fluorescence light is imaged onto a photomultiplier (Hamamatsu R5600-U-01) by a set of lenses, the optical axes of which lies perpendicularly to both the laser and the
cluster beam. A second photomultiplier monitors background photons (laser stray light and LIF of PTCDA molecules ejected effusively from the oven, i.e.~molecules not attached to a cluster). The laser system used for the cluster measurements consists of a pulsed dye laser (Sirah Cobra) pumped by a Nd:YAG laser (Edgewave IS-IIIE) at a repetition rate of 1\,kHz and pulse lengths of about 10\,ns. The dye laser system has a maximal energy of about 100\,$\mu$J per pulse but was attenuated to about 1\,$\mu$J per pulse in order to suppress saturation effects.

\section{Results}
\subsection{PTCDA embedded in neon matrix}
\begin{figure*}
{\center\includegraphics[width=14cm]{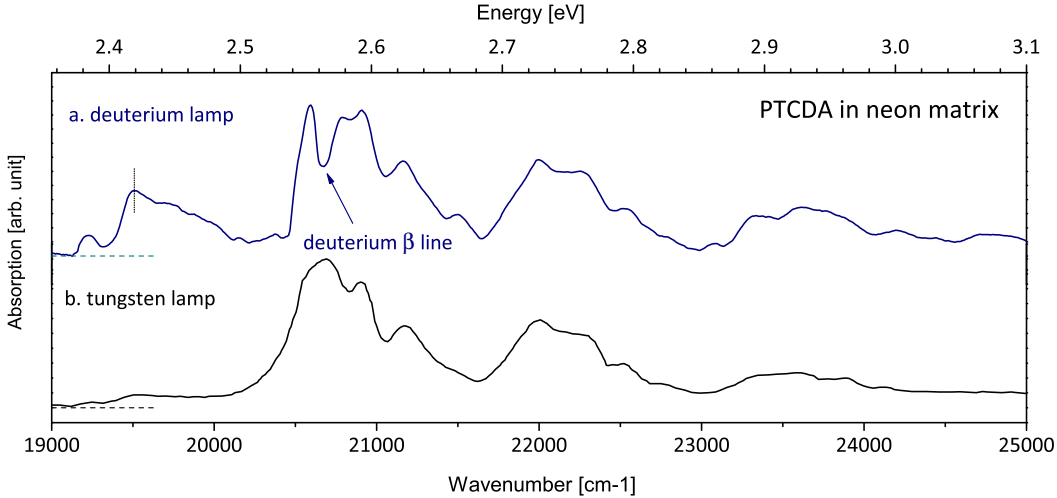}}
  \caption{Absorption spectra of PTCDA in a neon matrix excited with a deuterium lamp (a) and a tungsten lamp (b). In (a), the peak at around 19\,500\,cm$^{-1}$~is attributed to charged PTCDA molecules and the depletion at around 20\,600\,cm$^{-1}$~is a side effect of the presence of the Balmer line D$\beta$ at this wavenumber.}
  \label{lampComparison}
\end{figure*}
The absorption spectra of PTCDA embedded in a neon matrix have been recorded with the QTH and D$_2$ lamps as shown in Fig.~\ref{lampComparison}. The spectra are characterized by four broad bands (FWHM 500--700\,cm$^{-1}$) containing sub-structures. However, when comparing the two spectra, some discrepancies are observed which result from peculiar distributions of emission frequencies of these lamps. The tungsten lamp has an emission spectrum limited to the visible range (roughly between 12\,500\,cm$^{-1}$~and 25\,000\,cm$^{-1}$), whereas the deuterium lamp emits further into the UV/VUV (the continuous part of the spectrum extends to about 55\,000\,cm$^{-1}$). The VUV contribution of the D$_2$ lamp  has direct implications on the absorption spectra of the PTCDA-doped matrix. The spectrum (Fig.~\ref{lampComparison}\,a) shows an additional band at about 19\,500\,cm$^{-1}$, which is hardly observable in the spectrum measured with the tungsten lamp (Fig.~\ref{lampComparison}\,b). We assign this peak to PTCDA 
cations. The first ionization threshold of PTCDA is between 50\,000\,cm$^{-1}$~and 55\,000\,cm$^{-1}$,\cite{Karl1992,Hirose1994,Hill1998} which can be accessed in a 1-photon process with the D$_2$ lamp but not with the QTH lamp. This interpretation is corroborated by different studies on neutral and charged PAH molecules embedded in rare gas matrices showing that PAH ions absorb in the visible-near IR domain, with typical red-shifts of some hundreds or some thousands of wavenumbers compared to the absorption bands of the neutral molecules.\cite{Salama1995} According to our measured spectra the ions absorb about 1\,200\,cm$^{-1}$~red-shifted, a value which is close to previous measurements on similar molecules. E.g., for pentacene in neon a red-shift of about 1\,200\,cm$^{-1}$~has been observed of the absorption bands of cations compared to neutral molecules.\cite{Halasinski2000} For perylene, similar red-shifts have been observed for anions ($\sim900$\,cm$^{-1}$) and for cations ($\sim$\,1\,700\,cm$^{-1}$).\
cite{Halasinski2003} The second obvious difference of both spectra is the pronounced dip at $\sim$\,20\,600\,cm$^{-1}$~which is related to the $\beta$ Balmer line of deuterium. The high lamp intensity at this wavenumber results in a saturation of the absorption.

The inhomogeneous broadening effects like e.g.~the presence of different site isomers of PTCDA or matrix defects hamper a precise assignment of the spectral features and their substructures. The depositing of PTCDA at the surface of rare gas matrices would decrease inhomogeneous broadening effects. However, in the present case this is not achievable for technical reasons. As an alternative, a beam of large neon clusters is used. After the formation of the clusters, single PTCDA molecules are attached. If the clusters are and, upon doping, remain solid, the chromophore is expected to reside on the cluster surface and hence neon clusters provide a weaker interacting environment (smaller spectral shift and broadening) compared to bulk matrix isolation. On the contrary, if neon clusters are liquid-like, the chromophore is expected to immerse, resulting in similar conformations as in bulk matrices and therefore leading to a similar spectral response.
\begin{figure}
{\center\includegraphics[width=8.5cm]{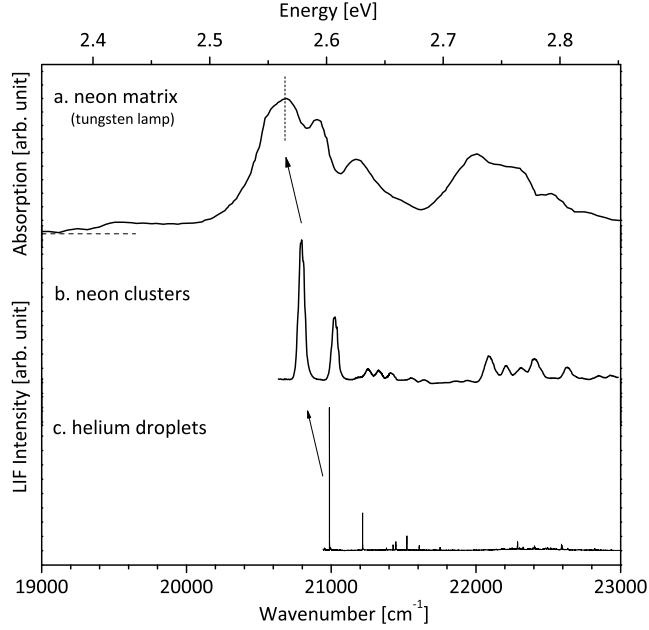}}
  \caption{Absorption spectrum of PTCDA in neon matrix measured with a tungsten halogen lamp (a), compared to the LIF excitation spectrum of PTCDA
molecules attached to large neon clusters, N=6\,000 (b) or embedded in helium droplets (c).}
  \label{neonComparison}
\end{figure}
The excitation spectrum obtained for PTCDA-doped neon clusters is displayed in Fig.~\ref{neonComparison}\,b, and compared to the absorption spectrum in Ne (a) and the excitation spectrum in helium droplets (c). An evident correspondence between the neon cluster spectrum and the helium droplet spectrum is observed: the LIF spectrum of PTCDA-doped neon clusters shows the same vibrational modes as observed in helium droplets. The cluster spectrum is broadened and shifted to the red by about 190\,cm$^{-1}$~when compared to the helium droplet spectrum. The neon matrix absorption spectrum is red-shifted by about 300\,cm$^{-1}$~and its main bands as well as the main subfeatures, although much broader, can be clearly assigned to those of the cluster spectrum. The neon matrix is grown slowly at elevated temperatures leading to rather large grain sizes of the polycrystalline solid which is reflected in an enhanced optical transmission. Embedding molecules inside a Ne cluster one would expect a very similar environment 
as in the matrix case. The apparent differences, however, clearly point to the fact that PTCDA remains on the surface of the clusters which are solid for such sizes ($N=6\,000$). The quantitative agreement of vibrational modes will be discussed in the following section.

\subsection{PTCDA attached to large argon, neon and para-hydrogen clusters}\label{LIFmeasurement}
\subsubsection{Assignment of LIF absorption spectra}
\begin{figure}
{\center\includegraphics[width=8.5cm]{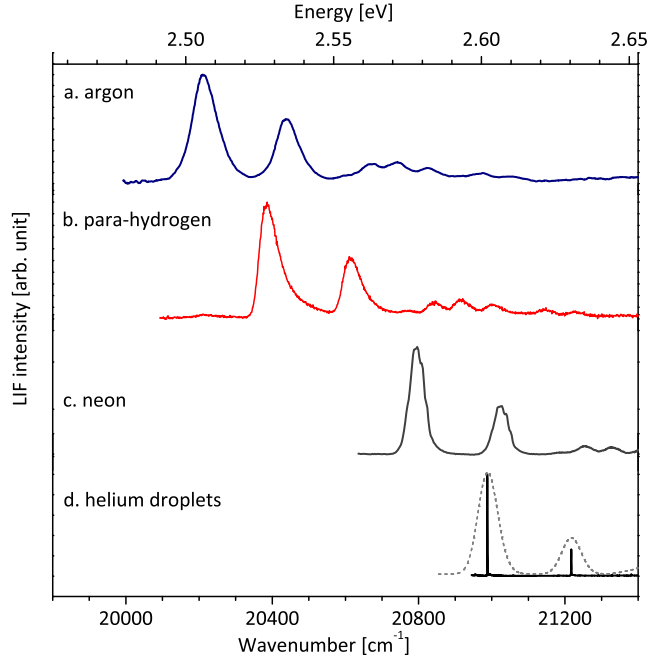}}
  \caption{LIF excitation spectra of PTCDA attached to large argon, para-hydrogen and neon clusters, compared to the spectra of PTCDA embedded in helium droplets.}
  \label{clusterLIF}
\end{figure}
Fig.~\ref{clusterLIF}~a-c compares the excitation spectra of single PTCDA molecules attached to large argon, para-hydrogen and neon clusters, to spectra of molecules attached to helium nano\-drop\-lets (Fig.~\ref{clusterLIF}~d.). In all three cases, vibronic bands are observed even though the resolution is 1-2 orders of magnitude poorer. The argon cluster spectrum of PTCDA is shifted to the red by 775\,cm$^{-1}$~compared to the one obtained in helium droplets. All vibronic bands observed have an identical FWHM of about 80\,cm$^{-1}$. The overall structure of the spectrum is similar to the one measured in helium droplets, no new lines are observed and the relative positions of the vibronic bands are identical in both cases.

\begin{figure}
\center\includegraphics[width=8.5cm]{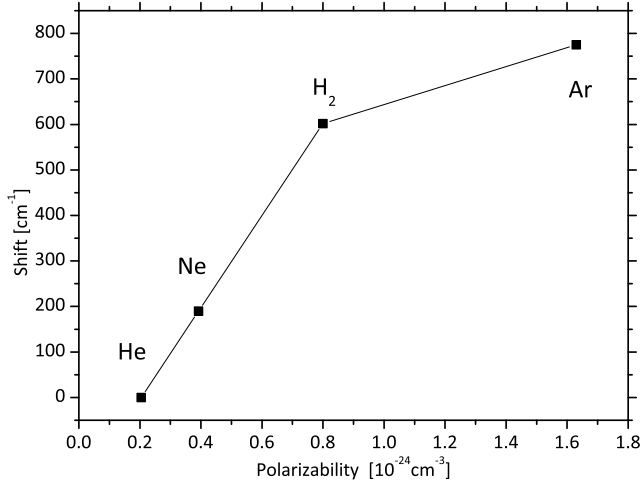}
  \caption{Shift of lines plotted versus the static polarizabilities of the cluster constituents taken from ref. \cite{lide2006crc}.}
  \label{polarizable}
\end{figure}

The same observations hold for para-hydrogen and neon clusters. However, due to the weaker interaction, red-shift and broadening are slightly reduced: 602\,cm$^{-1}$~shift for para-hydrogen (FWHM = 62\,cm$^{-1}$) and 190\,cm$^{-1}$~shift for neon (FWHM = 47\,cm$^{-1}$). If one compares this with the binding energy of the constituents within the cluster: Ar\,-\, 450\,cm$^{-1}$, H$_2$\,-\,50\,cm$^{-1}$~and Ne\,-\,120\,cm$^{-1}$,\cite{kuma2011} one finds that hydrogen, as a molecular and not rare gas constituent, does not follow the order in the amount of the observed shifts. However, when just taking in to account polarizabilities, the larger shift induced by the hydrogen clusters is understandable. In Fig.~\ref{polarizable} the shifts are plotted versus atomic/molecular polarizabilities. Although the order is right, we do not find the almost linear dependence, as has been observed before for bulk rare gas matrices and related organic dopants.\cite{Biktchantaev2002}

The cluster size dependence of the spectra has been studied using different expansion conditions. In the range covered (hundreds to thousands of atoms) no such dependence was found meaning that the cluster size does not influence the spectral signature of the chromophore. This implies that the structure of the surface is independent of the cluster size and effects on the curvature of the surface are negligible.

\begin{figure}
\center\includegraphics[width=8.5cm]{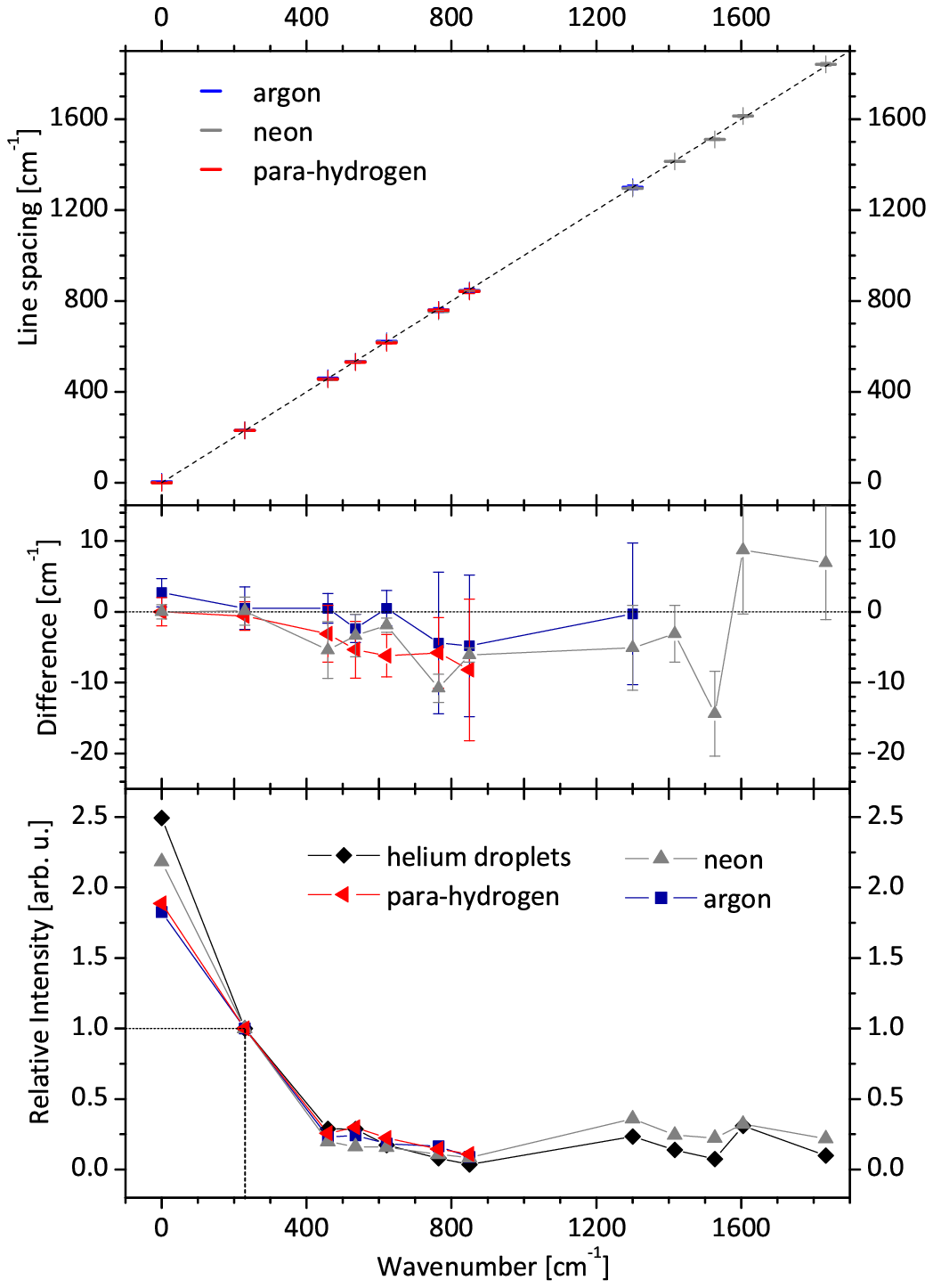}
  \caption{Extracted vibrational spacings relative to the 0-0 transition of PTCDA-doped argon, neon and para-hydrogen clusters compared to the corresponding lines in helium droplets (top). Difference in absolute numbers (middle). Comparison of relative intensities of the bands (bottom).}
  \label{compa_lines}
\end{figure}
\begin{center}
\begin{table*}
\caption{Peak positions and associated vibrational intervals for the $S_1 \rightarrow S_0$ transition of PTCDA molecules attached to large argon, neon and para-hydrogen clusters compared to helium droplets data.}
\begin{tabular}{p{1.5cm} p{2cm} p{1.5cm} p{2cm} p{1.5cm} p{2cm} c}
\hline
\hline
\multicolumn{6}{c}{rare gas and para-H$_2$ clusters} & helium \\
\cline{1-6}
\multicolumn{2}{c}{argon}& \multicolumn{2}{c}{neon} & \multicolumn{2}{c}{para-H$_2$} & droplets~\cite{W_mono}\\
\cline{1-6}
$\nu$ (cm$^{-1}$) & $\Delta\nu$ (cm$^{-1}$) & $\nu$ (cm$^{-1}$) & $\Delta\nu$ (cm$^{-1}$) &$\nu$ (cm$^{-1}$) & $\Delta\nu$ (cm$^{-1}$) & $\Delta\nu$
(cm$^{-1}$)\\
\hline
20212 & 0 & 20797.5 & 0 & 20386.4 & 0 & 0\\
20439 & 227 & 21027 & 229.5 & 20615 & 229 & 229.5\\
20672 & 460 & 21251 & 453.5 & 20842 & 456 & 459\\
20743 & 531 & 21329 & 532 & 20916 & 530 & 535.25\\
20827 & 615 & 21416 & 619 & 21001 & 615 & 620.8\\
20973 & 761 & 21551 & 753.5 & 21145 & 759 & 764.4 \\
-- & -- & 21641 & 844 & 21228 & 842 & 849.8\\
21511 & 1299 & 22093 & 1295.5 & -- & -- & 1300.3\\
-- & -- & 22211 & 1413.5 & -- & -- & 1417\\
-- & -- & 22308 & 1510.5 & -- & -- & 1532.2\\
-- & -- & 22411 & 1614 & -- & -- & 1603.5\\
-- & -- & 22638 & 1840.5 & -- & -- & 1832.6\\
\hline
\hline\label{T_largeclust}
\end{tabular}
\end{table*}
\end{center}
Fig.~\ref{compa_lines} (top) plots the relative position of each vibronic band measured in PTCDA-doped rare gas and hydrogen clusters with respect to the corresponding lines observed in helium droplets. The difference shown in absolute numbers (middle) does not point to a significant systematic change of the vibrational spacings when having the PTCDA molecule in the different environments. The relative intensities of the main vibronic lines are presented in Fig.~\ref{compa_lines} (bottom). As all the peaks of a given spectrum show identical shape and width, the maximal peak
intensities can be used for comparisons. The first vibrational mode (at about 230\,cm$^{-1}$) has been used for normalization because this mode is less prone to saturation than the 0-0 transition. The relative intensities follow an identical pattern for all the measured spectra. Only weak discrepancies concerning the 0-0 line are observed, attributed to saturation effects. Table \ref{T_largeclust} summarizes the extracted line positions.

Measurement in argon matrices could not be achieved successfully. Argon matrices are known to be polycrystalline with much smaller grain sizes than neon leading to a stronger dispersion of the light.\cite{Klein1976} In the present spectroscopic configuration the matrix is opaque. With the present experimental setup the study of para-hydrogen matrix is not possible. Nevertheless, the analogy of the measurements in neon allow to draw conclusions for argon and para-hydrogen as well. As outlined in the introduction, for the cluster sizes considered in the present work, neon, argon, and para-hydrogen clusters are solid. As discussed before, in the case of neon the pronounced difference between the matrix absorption spectrum and the cluster LIF excitation spectrum clearly demonstrates that the PTCDA molecule resides on the surface of solid neon clusters. Moreover, for argon and hydrogen clusters one also observes a quite similar behavior of broadening and shifting of lines. The relative positions of vibronic bands 
are identical to the values measured in helium droplets. If the PTCDA molecules were located inside such solid clusters which have an even enhanced interaction, one would clearly expect a much broader and less resolved spectrum. The surface location is further confirmed in recent results on
complexes made of a single PTCDA molecules and argon or para-hydrogen clusters embedded in helium droplets.\cite{DvorakSC} In that case the use of helium droplets allows for the generation of surface-located but also of embedded PTCDA molecules. The measured spectra clearly show both for argon and for hydrogen a larger shift of embedded molecules. Finally a surface location is corroborated by previous measurements involving large argon clusters doped with barium or calcium atoms by a pickup process where only surface sites were observed. This was interpreted as proof of the solid phase for such clusters.\cite{visticot1994,Gaveau2002} In conclusion our work confidently shows that large rare gas and para-hydrogen clusters made of hundreds to thousands of atoms are solid and doping with PAH molecules results in surface site occupation only.

\subsubsection{Line Broadening and line shape}
\begin{figure*}
{\center\includegraphics[width=14cm]{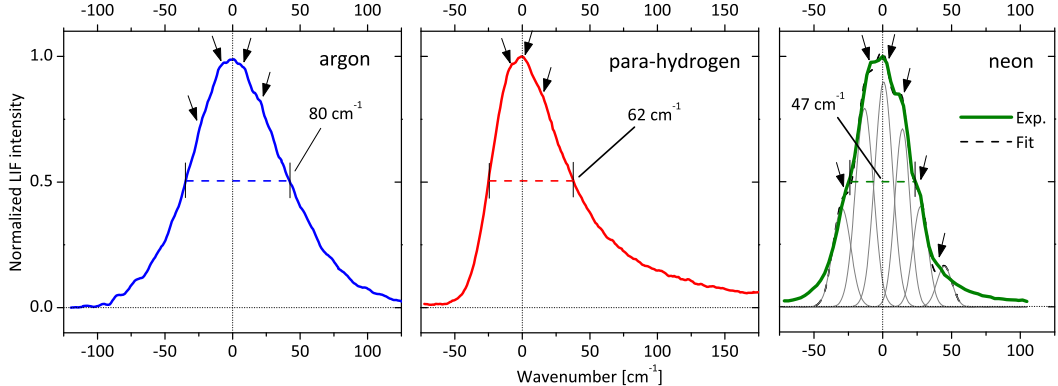}}
  \caption{Line shapes of the $S_1 \leftarrow S_0$ transition peak of PTCDA molecules attached to argon, para-hydrogen and neon clusters. Different subfeatures (arrows) are observed which are attributed to different site isomers}
  \label{peak}
\end{figure*}
Besides the red-shift, the cluster species has a strong impact on the line broadening as well as on the line shape. As for the global spectral shift, these parameters do not show a dependence on the cluster size in the present size range.
The shape of the 0-0 transition of PTCDA attached to argon, para-hydrogen and neon clusters is displayed in Fig.~\ref{peak}. We exemplified discuss the 0-0 peaks as all the vibrational lines show the same features. Quite some experiments and theory involving much smaller clusters (up to a few tens of atoms) have been made in the past, most of which dealing with argon. Tetracene-Ar$_N$ complexes were shown to have a monotonically increasing linewidth converging for $N>20$ atoms to a few tens of cm$^{-1}$.\cite{BenHorin1992}  This is interpreted as a manifestation of an inhomogeneous broadening related to the overlap of many coexisting structural isomers which are all homogeneously
broadened. Moreover, the clusters' temperature, which is increasing with the cluster size up to some hundreds of atoms,\cite{rytkonen98} induces some broadening. E.g.~the typical FWHM of the calculated spectrum of tetracene-Ar$_{26}$ increases from 33 to 72\,cm$^{-1}$~for a cluster temperature rising from 25 to 35\,K.\cite{BenHorin1992}

In general, our measured lines all show an asymmetry with a more or less pronounced tail extending to the blue. The asymmetry is rather weak for argon and neon but notably stronger when using para-hydrogen clusters. In the past, calculations have been made on similar systems, perylene-Ar$_N$ complexes, with the semiclassical spectral density method.\cite{Heidenreich1993} This model has been developed by Fried \textit{et al.}~\cite{Fried1992} and takes into account the diffusion of the atoms, the fluctuation of the bond length and of the electronic energy gap, the
fluctuations of the solvent as well as the cluster asymmetry. Applied to perylene- or benzene-argon complexes this method proved to satisfactorily reproduce experimental measurements and also explain the asymmetry toward higher energy of the electronic transitions of the chromophore. However, the calculated FWHM strongly decreases with decreasing temperature and is for temperatures below 20K quite smaller than found in the present measurements. One reason might be the much smaller cluster sizes considered. For calculated very small clusters ($N<5$), isomers having different shifts appear as separated lines. In the present work, substructures are observed (designated by arrows in Fig.~\ref{peak}). Their presence and position are again independent of the cluster size or any external parameters (laser intensity, doping parameters, etc.). In the spectra presented in Fig.~\ref{peak}, these subfeatures are particularly visible for neon clusters. The presence of such subfeatures clearly hints to inhomogeneously 
broadening due to different chromophore-cluster conformations (the so-called site-isomers). In the case of neon clusters, the six different subfeatures are fitted with
six Gaussian distributions (all of them having here an identical FWHM of about 12\,cm$^{-1}$), shifted one to each other. Such an interpretation of a discrete number of isomers having specific spectra can only be true when having well-defined geometric structures (FCC, icosahedral) which again confirms the solid nature of the studied clusters. It also would mean that the deposited molecule anneal to a well defined number of site isomers.

\subsubsection{Formation of dimers and complexes}
The overall intensity of the fluorescence signal of PTCDA attached to argon, neon and para-hydrogen clusters is distinctly higher than for helium droplets (4-5 times for neon clusters). Several reasons can in principle contribute to that. One could, e.g., have a higher cluster density in the case of argon, neon or para-hydrogen clusters compared to helium droplets.  On the other hand, these large solid clusters are able to capture more than one PTCDA molecule. However this does not result in the formation of PTCDA dimers or oligomers and the spectroscopic identity is conserved.
In this way multiple absorbers per cluster are provided which results in a higher total signal intensity. In helium droplets the multiple doping of PTCDA gives rise to the formation of oligomers, whose spectral signature is quite different from that of the monomer.

The number of picked-up molecules can be determined when studying the pickup statistics dependent on the density of molecules in the doping cell. For clusters as well as for droplets, the pick-up process is governed by a Poissonian statistics.\cite{Buenermann2011} If one neglects the shrinking of the clusters upon the doping and therefore a reduced capture cross-section, the probability $P(N)$ to collect $N$ particles after crossing the pick up volume can be written as:
\begin{equation}
P(N)=\frac{z^N}{N!}e^{-z}~,
\end{equation}
where $z$ is the mean number of doping particles picked up by each cluster. This parameter is proportional to the partial pressure of the doping particle in the oven $p_{dop}$ and is given as follow:
\begin{equation}
z=\frac{\sigma_{cap} L}{k_B T}\sqrt{\frac{\langle v_{Clus}^2\rangle + \langle v_{dop}^2\rangle}{\langle v_{Clus}^2\rangle}}\;p_{dop}~,
\end{equation}
where $L$ is the length of the pickup region, $\sigma_{cap}$ the capture cross-section of the cluster which we assume to be proportional to $n_0^{2/3}$, $n_0$ being the mean cluster size. $\langle v_{Clus}^2\rangle$ and $\langle v_{dop}^2\rangle$ are the mean square velocities of the clusters and $T$ the temperature of the doping cell. A careful study of the Poissonian distribution for low densities in the doping cell gives valuable information on the doping process and allows the distinction between measured signals coming from a single pickup (linear slope), the pickup
of two molecules (quadratic slope) or multiple doping (higher polynomial slopes).

\begin{figure}
{\center\includegraphics[width=8.5cm]{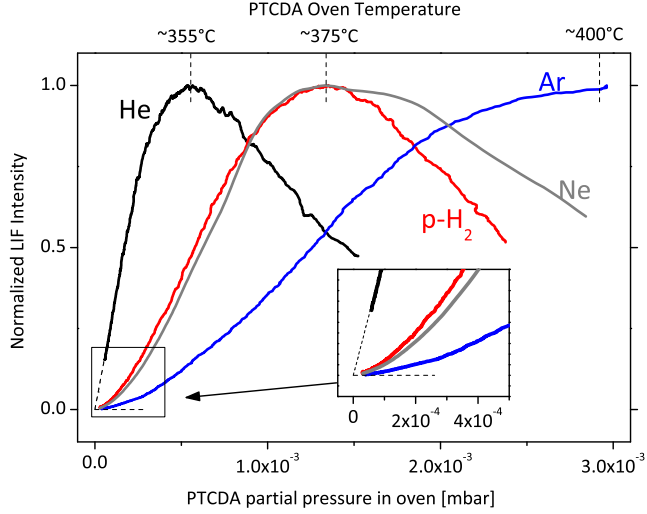}}
  \caption{LIF intensity of the $S_1 \leftarrow S_0$ transition of PTCDA molecules attached to helium droplets, argon, neon and para-hydrogen
clusters as a function of the PTCDA partial pressure in the oven.}
  \label{poisson}
\end{figure}

When recording LIF intensity as a function of the partial pressure of the doping particle in the oven one can assign the contributions of multiple doped clusters to the spectrum. In Fig.~\ref{poisson}, the intensity of the 0-0 line of PTCDA monomer attached to helium droplets or argon, neon or para-hydrogen clusters are displayed as a function of the PTCDA partial pressure in the oven.
The maxima of these distributions are reached at higher temperatures for hydrogen, neon (both at $\sim$375\,°C) and argon clusters ($\sim$400\,°C) than for helium droplets ($\sim$355\,°C). Having higher intensities of monomer absorptions at higher doping levels intuitively indicates that signals come from multiple doped clusters. However, since the position of the maximum is dependent on the pickup cross-section, there is quite some dependence on the cluster size used in the experiment and on a possible different sticking coefficient. In the present case the argon, neon and para-hydrogen
clusters have a mean size of 450, 6\,000 and 14\,000 atoms/molecules, respectively. Using general geometric considerations applied to a FCC packing for these clusters, one obtains mean values for the cluster diameter of about 60\,\AA, 125\,\AA~and 200\,\AA, respectively. For comparison, the helium droplets  have a mean size of roughly 20\,000 atoms which corresponds to a diameter of about 100\,\AA. From these numbers one would not expect all the maxima lying beyond the position of the maximum of helium droplets. Furthermore, the most interesting observation is that for hydrogen, neon and
argon, the onset of the curves are not linear but clearly dominated by higher order polynomials. This unambiguously shows, that the spectra come from multiply doped clusters, all contributing independently to the monomer spectral signature. Since no molecular complexes are formed which would alter the optical response, mobility of the PTCDA molecules on the cluster surface must be strongly quenched. This is in contradiction to previous measurements involving smaller chromophores. It has been shown that barium atoms can move freely on argon clusters to form dimers and that such clusters can work successfully as reaction centers for chemical reactions.\cite{Lallement1993a,Biquard1994} Our results clearly show that this apparently does not apply for larger PAH molecules such as PTCDA where the binding energy is presumably much larger.

\section{Conclusion}
We present the excitation spectra of PTCDA attached to large argon, neon and para-hydrogen clusters made of thousands of atoms/molecules. LIF excitation spectra are compared to absorption spectra in solid Ne matrices. The similarity of these spectra to spectra obtained in helium droplets allows the assignment of vibrational modes of PTCDA molecules. The comparison of the excitation spectra with the absorption spectrum of PTCDA in a neon matrix as well as the presence of substructures in the spectral lines prove that only surface sites are occupied by PTCDA and that the cluster
surface is solid. Moreover, the precise characterization of the doping process of these clusters points out that the mobility of the chromophore on the cluster surface is strongly suppressed, preventing the formation of PTCDA dimers or oligomers, in contradiction to previous measurements where atoms and smaller molecules were used.

\section{Acknowledgements}
Fruitful discussions with Takamasa Momose and Andrey Vilesov are gratefully acknowledged.


%


\end{document}